\documentclass[twocolumn,aps,prx]{revtex4-1}

\usepackage{amsmath}
\usepackage{amsfonts}
\usepackage{amssymb}
\usepackage[colorlinks=true,citecolor=blue,linkcolor=red]{hyperref}
\usepackage{graphicx}
\usepackage{bbold}					
\usepackage[dvipsnames]{xcolor}

\usepackage{array,booktabs,ragged2e}	
\usepackage{soul}						
\usepackage{cancel}						
\usepackage{tabularx}					
\usepackage{multirow}
\usepackage{hhline}
\usepackage{adjustbox}
\usepackage{sidecap}

\usepackage[utf8]{inputenc}
\usepackage[T1]{fontenc}

\newcolumntype{P}[1]{>{\raggedleft\arraybackslash}p{#1}}
\newcolumntype{R}[1]{>{\centering\arraybackslash}p{#1}}

\begin{document}


\title{Magnetoresistance from Fermi Surface Topology}

\author{ShengNan Zhang$^{1,2}$}
\author{QuanSheng Wu$^{1,2}$}
\email{quansheng.wu@epfl.ch}
\author{Yi Liu$^{3}$}
\author{Oleg V. Yazyev$^{1,2}$}
\email{oleg.yazyev@epfl.ch}

\affiliation{$^{1}$Institute of Physics, Ecole Polytechnique F\'{e}d\'{e}rale de Lausanne (EPFL), CH-1015 Lausanne, Switzerland}
\affiliation{${^{2}}$National Centre for Computational Design and Discovery of Novel Materials MARVEL, Ecole Polytechnique F\'{e}d\'{e}rale de Lausanne (EPFL), CH-1015 Lausanne, Switzerland}
\affiliation{${^{3}}$The Center for Advanced Quantum Studies and Department of Physics, Beijing Normal University, 100875 Beijing, China}
\date{\today}

\begin{abstract}
Extremely large non-saturating magnetoresistance has recently been reported for a large number of both topologically trivial and non-trivial materials. Different mechanisms have been proposed to explain the observed magnetotransport properties, yet without arriving to definitive conclusions or portraying a global picture. In this work, we investigate the transverse magnetoresistance of materials by combining the Fermi surfaces calculated from first principles with the Boltzmann transport theory approach relying on the semiclassical model and the relaxation time approximation. We first consider a series of simple model Fermi surfaces to provide a didactic introduction into the charge-carrier compensation and open-orbit mechanisms leading to non-saturating magnetoresistance. We then address in detail magnetotransport in three representative materials: (i) copper, a prototypical nearly free-electron metal characterized by the open Fermi surface that results in an intricate angular magnetoresistance,
(ii) bismuth, a topologically trivial semimetal in which very large magnetoresistance is known to result from charge-carrier compensation, and (iii) tungsten diphosphide WP$_2$, a recently discovered type-II Weyl semimetal that holds the record of magnetoresistance in compounds. In all three cases our calculations show excellent agreement with both the field dependence of magnetoresistance and its anisotropy measured at low temperatures. Furthermore, the calculations allow for a full interpretation of the observed features in terms of the Fermi surface topology. Our study thus establishes guidelines to clarifying the physical mechanisms underlying the magnetoransport properties in a broad range of materials. These results will help addressing a number of outstanding questions, such as the role of the topological phase in the pronounced large non-saturating magnetoresistance observed in topological materials.
\end{abstract} 

\maketitle

\section{Introduction}


Magnetoresistance (MR) is the change of electrical resistance in an applied magnetic field. Magnetoresistance is commonly defined as MR$(B) = [\rho(B)-\rho(0)]/\rho(0)$, where $\rho(0)$ and $\rho(B)$ are electrical resistivities in zero field and in applied magnetic field $B$, respectively. Positive MR typically occurs in metals, semiconductors, and semimetals, while negative MR is seen in magnetic materials. MR is a relatively weak effect in most non-magnetic compounds, being characterized by quadratic field dependence in low fields that saturates to a magnitude of a few percent in case of metals. In contrast, giant MR~\cite{GM1-PRL.61.2472, GM2-PRB.39.4828} and colossal MR~\cite{CM1-Ramirez1997, CM2} occur in multilayer composed of magnetic and non-magnetic layers and in manganese-based perovskite oxides, respectively, exhibiting values up to several orders of magnitude. Materials with large MR found applications in magnetic devices for data storage~\cite{Magneticdevice-Lenz1990, Magneticmemory-Moritomo1996, Magneticdevice-DAUGHTON1999334}, which has been stimulating fundamental and applied research into this transport phenomenon. 

More recently, large MR effects at low temperature, distinct from giant and colossal MR, have been reported for numerous materials many of which host topological electronic phases. Dirac semimetals such as graphene~\cite{Friedman2010,graphene-Gopinadhan2015}, Cd$_3$As$_2$~\cite{Cd3As2-Liang2014, Cd2As32-PRX.5.031037,PhysRevB.92.081306}, and Weyl semimetals belonging to the TaAs family~\cite{TaAs-PRX.5.031023} show linear field-dependent MR. The latter was argued to result from either a quantum effect near the crossing point of the linear valence and conduction bands in magnetic fields exceeding the quantum limit~\cite{QM-PRB.58.2788}, or mobility fluctuations caused by disorder~\cite{Cd2As33-PRL.114.117201,1367-2630-18-5-053039}.  
Type-II Weyl semimetals~\cite{Soluyanov2015} WTe$_2$~\cite{MRWTe2-Ali2014}, MoTe$_2$~\cite{MoTe2-PRB.94.235154,doi:10.1063/1.4995951} and WP$_2$~\cite{WP2-Kumar2017,PhysRevB.96.121107,PhysRevB.96.121108} show nearly quadratic field dependence of MR. Different mechanisms responsible for the extremely large MR, often referred to as XMR, in these non-magnetic materials have been suggested. In addition to topological protection possibly playing an important role, the classical two-band model predicts quadratic field dependence in compensated semimetals, in which the density of electrons equals to that of holes. Any difference in charge-carrier densities leads to deviation from the quadratic field dependence, and eventual saturation of MR. The observed extremely large  non-saturating MR in WTe$_2$, claimed to be caused by almost perfect compensation, shows nearly parabolic field dependence of MR~\cite{MRWTe2-Ali2014}.

Meanwhile, many topologically trivial materials also show XMR, with bismuth being the most well-known example~\cite{bismuth-PhysRev.91.1060, bismuth-PRX.5.021022,0953-8984-30-31-313001}, but also including PdCoO$_2$~\cite{PdCoO2-PRL.111.056601}, PtSn$_4$~\cite{PtSn4-PRB.85.035135}, NbSb$_2$~\cite{NbSb-Wang2014}, LaSb~\cite{LaSb-FallahTaftiE3475}, YSb~\cite{YSb-PRL.117.267201,PhysRevB.96.075159},  MoAs$_2$~\cite{PhysRevB.96.241106}, NbAs$_2$~\cite{TaAs2NbAs-PRB.93.184405,PhysRevB.93.195119}, TaAs$_2$~\cite{TaAs2NbAs-PRB.93.184405}, TmSb~\cite{TmSb-PRB.97.085137}, the centrosymmetric $\alpha$-phase of WP$_2$~\cite{PhysRevB.97.245101,PhysRevB.97.245151} and others. Topological protection is irrelevant in these materials, hence two mechanisms related to the Fermi surface topology have been considered to explain XMR in these topologically trivial materials. One mechanism is the compensation between electron and hole charge carriers~\cite{chambersbook, pippardbook}, as in the case of WTe$_2$ mentioned above. Nonetheless, in materials like YSb~\cite{YSb-PRL.117.267201}, a quantitative analysis gives an electron-hole concentration ratio of $\approx 0.81$, departing far from the perfect compensation in the classical isotropic two-band model. While an unequal mobility of electron and hole carriers was suggested to explain this discrepancy, the two-band model was still insufficient to provide a consistent description of all available experiment data.
The other mechanism is possible in the case of non-closed Fermi surfaces~\cite{chambersbook, pippardbook}, which results in open-orbit trajectories of charge carriers driven by the Lorentz force under magnetic field. Delafossite PdCoO$_2$ was found to display a large MR, reaching 10$^5$\% at 2~K temperature and 14~T magnetic field, for electric current along the interlayer axis, which is due to the motion of carriers along the open orbits according to the experimental and theoretical work by Takatsu and co-workers~\cite{PdCoO2-PRL.111.056601}.

Although the family of materials showing experimentally confirmed large MR keeps expanding, no consistent general theory explaining this phenomenon for a broad range of cases has been developed so far. Intuitively, the Fermi surface is playing an important role since within the semiclassical approximation its topology translates into the trajectory of charge carriers, and hence electrical conductivities, under applied magnetic field. Since last century a large number of models of magnetotransport relying on Fermi surfaces approximated by simple geometric shapes, such as cubes and spheroids, have been introduced. The most successful example is the model of magnetotransport in bismuth. Abeles and Meiboom~\cite{bismuth-theory-PhysRev.101.544}, Aubrey~\cite{Aubrey-0305-4608-1-4-321} constructed accurate models of magnetoconductivity of bismuth  approximating its Fermi surface by ellipsoids appropriately arranged in momentum space. However, the Fermi surfaces differ from one material to another, and can be very complex, therefore a universal theory relying on the direct introduction of the Fermi surfaces is required for studying magnetotransport phenomena in a broad range of materials.

In this work, we present a systematic study of transverse MR by using the Botlzmann transport theory~\cite{chambersbook, pippardbook} within the relaxation time approximation. 
With the help of models we first demonstrate in a didactic manner how the compensation of charge carriers, open orbits and detailed geometry of the Fermi surface result in MR, in particular the non-saturating XMR. We then consider MR in three different representative materials -- copper, bismuth and WP$_2$ -- relying on the Fermi surfaces obtained from first-principles density functional theory (DFT) calculations. Surprisingly, in all cases the calculated MR as a function magnetic field orientation and strength show very good agreement with available experimental data even assuming constant relaxation times that do not depend on momentum and band index. For these seemingly unrelated materials, we find that the topology of the Fermi surface plays a crucial role in the magnetotransport properties, and our calculations allow for a complete interpretation of the observed features. Our work establishes a methodology that facilitates understanding of magnetotransport in a broad range of materials, including topological materials and other compounds displaying very high measured values of MR.

The three representative materials we have chosen for our detailed investigation can be described as follows.
Copper is a prototypical metal, in which the Fermi surface is only one level more complex than that of free electrons. The effect of the periodic potential is sufficiently strong to create Fermi surface ``necks'' touching the boundary of the Brillouin zone. This gives rise to a variety of Fermi surface cross-sections upon applying magnetic fields in different directions. These cross-sections result in both electron and hole closed orbits as well as in open orbits depending on the field orientation, hence resulting in anisotropic MR. Our calculations are able to reproduce all the delicate features of angular MR in very good agreement with experiments. 
Bismuth is a semimetal that exhibit XMR of $~1.6 \times 10^7\% $ at $T=4.2$~K in magnetic field of 5~T~\cite{bismuth-PRX.5.021022}, that results from compensation of electron and hole charge carriers. When current is applied along the three high-symmetry axes, MR shows very distinct anisotropic patterns caused by the multi-valley Fermi surface with three electron and one hole pockets. In that sense, the peculiar Fermi surface of bismuth makes itself a reference material for exploiting anisotropic MR phenomena. It is worth mentioning that the quantum limit in bismuth is reached at relatively low fields due to the low carrier density, hence our considerations within the semiclassical model apply only to the low-field regime.
Tungsten diphosphide WP$_2$ has recently been predicted to host the type-II Weyl semimetals phase robust against various perturbations~\cite{WP2-PRL.117.066402}. Subsequent experiments performed on this material revealed extremely high MR reaching $~4.2 \times 10^6\% $ at $T=2$~K in magnetic field of $9$~T, the largest reported in a compound, as well as a number of other intriguing properties such conductivity comparable to that of metals and a very high residual resistivity ratio~\cite{WP2-Kumar2017}. Unlike bismuth, however, WP$_2$ has a significantly more extended Fermi surface comprised of open hole and closed electron pockets, which are intuitively attributed to the measured anisotropic MR.
This makes WP$_2$ another interesting material from the point of view of magnetotransport that is still poorly understood. Our calculations show that the details of the the Fermi surface geometry are the key ingredient in explaining its highly anisotropic MR due to a novel charge-carrier compensation effect.

Our paper is organized as follows. In Section~\ref{Methodology} we present the details of our computational methodology. Section~\ref{model} considers magnetotransport in model systems. 
Section~\ref{Results} discusses the results for the  representative real materials. Finally, Section~\ref{Summary} summarizes our work.

\section{Methodology}
\label{Methodology}
The Boltzmann transport theory has been successfully used for explaining magnetotransport anisotropy observed in quasi-one-dimensional~\cite{1DMR1-PRL.72.3714, 1DMR2-JPSJ.65.3973} and quasi-two-dimensional materials~\cite{PdCoO2-PRL.111.056601}. Later, a numerical implementation of this approach using the Wannier-interpolated first-principles band structures was introduced for studying the magnetoconductivity of MgB$_2$~\cite{PhysRevLett.101.067001,liu-PRB.79.245123}. We have implemented this numerical approach based on the maximally localized Wannier functions tight-binding model~\cite{tbH-RMP.84.1419} that was constructed by using the Wannier90~\cite{MOSTOFI20142309} and WannierTools~\cite{WT-wu2017wanniertools} packages.

The conductivity tensor $\bold{\sigma}$ is calculated in presence of an applied magnetic field by solving the Boltzmann equation within the relaxation time approximation as~\cite{Neilbook}
\begin{equation}
\sigma^{(n)}_{ij}(\bold{B})=\frac{ e^2}{4 \pi^3} \int d\bold{k} \tau_n \bold{v}_n(\bold{k})  \bold{\bar{v}}_n(\bold{k}) 
\left(- \frac{\partial f}{\partial \varepsilon} \right)_{\varepsilon=\varepsilon_n(\bold{k})},
\label{eqn-sigmaij}
\end{equation}
where $e$ is the electron charge, $n$ is the band index, $\tau_n$ is the relaxation time of $n$th band that is assumed to be independent on the wavevector $\bold{k}$, $f$ is the Fermi-Dirac distribution, $\bold{v}_n(\bold{k})$ is the velocity defined by the gradient of band energy
\begin{equation}
\bold{v}_n(\bold{k})=\frac{1}{\hbar} \nabla_{\bold{k}} \varepsilon_n(\bold{k}),
\label{eqn-velocity}
\end{equation}
and $\bar{\bold{v}}_n(\bold{k})$ is the weighted average of velocity over the past history of the charge carrier
\begin{equation}
\bar{\bold{v}}_n(\bold{k}) = \int^0_{-\infty} \frac{dt}{\tau_n} e^{\frac{t}{\tau_n}} \bold{v}_n(\bold{k}(t)) .
\label{eqn-aver_velo}
\end{equation}
The orbital motion of charge carriers in applied magnetic field causes the time evolution of $\bold{k}_n(t)$, written as, 
\begin{equation}
\frac{d \bold{k}_n(t)}{dt} = - \frac{e}{\hbar} \bold{v}_n(\bold{k}(t)) \times \bold{B}
\label{eqn-evol_k}
\end{equation}
with $ \bold{k}_n(0)=\bold{k}$. The trajectory $\bold{k}(t)$ can be obtained by integrating Eq.~(\ref{eqn-evol_k}). As a consequence, $\bar{\bold{v}}_n(\bold{k})$ can be calculated as the weighted average of the velocities  along the trajectory $\bold{k}(t)$ according to Eq.~(\ref{eqn-aver_velo}). In this semiclassical picture, the Lorentz force does no work on charge carriers since it is perpendicular to $\bold{v}_n(\bold{k})$, and therefore energy $\varepsilon_n(\bold{k})$ remains constant as $\bold{k}$  evolves in time. This is also evident from Eq.~(\ref{eqn-velocity}), implying that $\bold{v}$ is normal to the constant energy surface, and consequently $\dot{\bold{k}}$ 
is tangential to it. Since $\dot{\bold{k}}$ is also perpendicular to magnetic field $\bold{B}$, it follows that the $\bold{k}$ vector traces out an orbit which is a cross-section of the Fermi surface by a plane normal to $\bold{B}$.

According to the mutual orientation of magnetic field and current, one distinguishes two types of MR -- transverse and longitudinal. Since charge carriers are acted upon by the Lorentz force in the field which direction is perpendicular to velocity, we consider only the transverse MR in this work. This is nevertheless sufficient to provide a very rich playground for the comparison with experimental results and for discussing the underlying mechanisms.

Furthermore, we assume the relaxation time approximation and neglect interband scattering events and magnetic breakdown. In most multi-band materials, the relaxation times of different bands are different, and are usually difficult to determine. 
In order to provide a more general view we plot the results of our calculations as a function of combined variable $B\tau$, which corresponds to a dimensionless quantity $\omega \tau =\frac{e B \tau}{m^*}$. The latter represents a complete revolution of the cyclotron orbit before a carrier is scattered, with $m^*$ being the cyclotron mass.
In the case of multi-band systems, such as the semimetals discussed in our work, the total conductivity is the sum of band-wise conductivities, that is then inverted resulting in the resistivity tensor $\hat{\rho} = \hat{\sigma}^{-1}$. In order to analyze the results of calculations for the investigated semimetal systems we will often plot the individual resistivities of electrons and holes.

\section{Model Fermi surfaces}
\label{model}

Before considering real materials, we shall first discuss several model scenarios that result in non-saturating and anisotropic MR. To begin with, we introduce a general form one-band Hamiltonian
\begin{equation}
H(\bold{k})=\epsilon_0 + t_a \cos(k_x a)+t_b \cos(k_y b)+t_c \cos(k_z c) .
\label{eqn-Hamiltonian}
\end{equation}
The case of the isotropic ($t_a = t_b = t_c$), that is free-electron, spherical Fermi surface exhibits no MR since the Lorentz force is compensated by the force due to the Hall voltage. Magnetoresistance appears once the Fermi surface becomes anisotropic. It grows proportional to $B^2$ in weak magnetic fields ($\omega \tau \ll 1 $), but saturates in strong fields ($\omega \tau \gg 1 $) as long as the Fermi surface is closed. The saturation value depends on how far the Fermi surface departs from the ideal free-electron spherical shape.

Although the anisotropy of the Fermi surface is not sufficient to cause non-saturating MR, as the following examples will show, the equality or near equality of electron and hole concentrations, commonly referred to as the charge-carrier compensation, provides an opportunity for non-saturating MR. Another mechanism giving rise to non-saturating MR involves open orbits resulting from the Fermi surfaces that are not closed. Below, we discuss non-saturating MR for models with compensated and open Fermi surfaces. 

\begin{figure}
\begin{center}
\includegraphics[width=8.5cm]{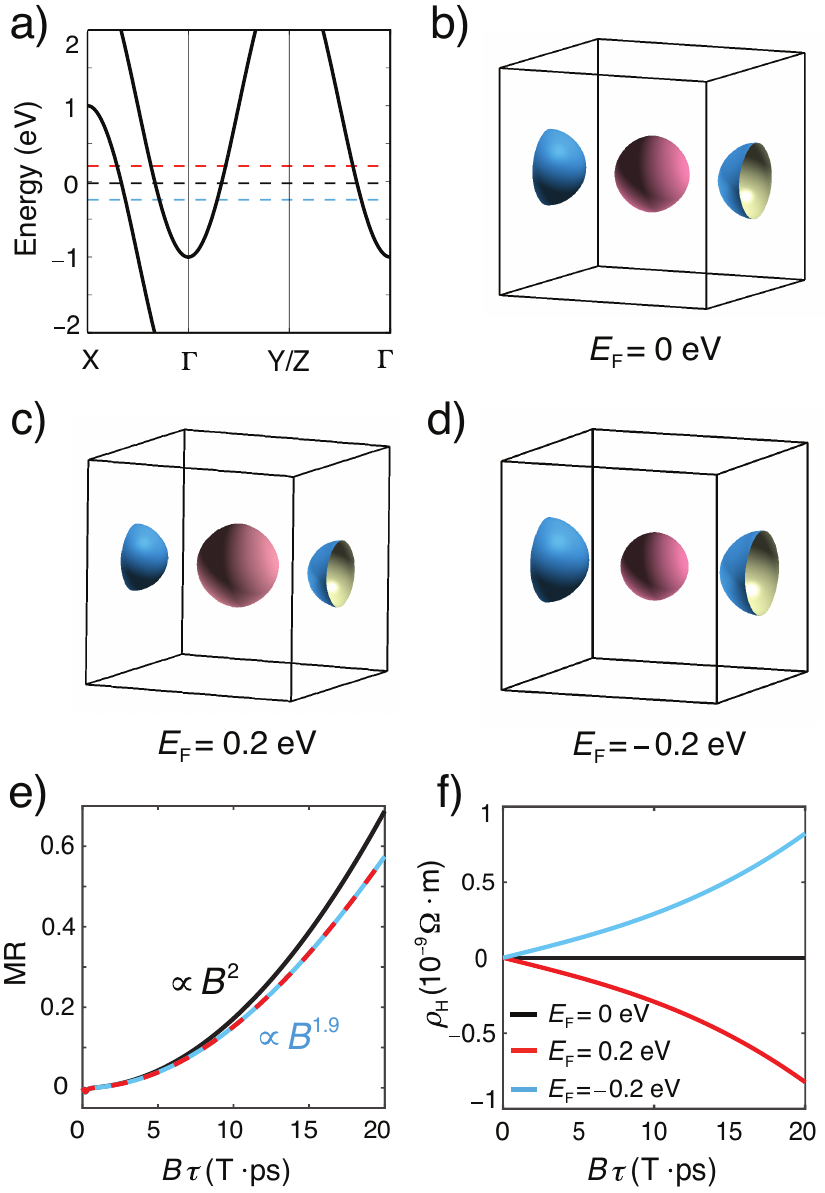}
\caption{(a) Band structure of the two-band model described by the Hamiltonian of Eq.~(\ref{2band_iso_H}). 
(b)-(d) Fermi surfaces for Fermi energies $E_\text{F} = 0$, $0.2$ and $-0.2$~eV, indicated in blue, black and red in panel (a), respectively.  Field dependence of (e) magnetoresistivity and (f) Hall resistivity for the three different Fermi energies.}
\label{fig1}
\end{center}
\end{figure}

\textit{Isotropic two-band model.} We start with a two-band Hamiltonian
\begin{equation}
H(\bold{k})=-2\cos(k_x) \sigma_0-[2\cos(k_y)+ 2\cos(k_z) - 5] \sigma_z
\label{2band_iso_H}
\end{equation}
where $\sigma_0$ is the $2 \times 2$ unit matrix and $\sigma_z$ is the corresponding Pauli matrix. The energy units are electronvolts. This Hamiltonian is comprised of electron and hole counterparts, and the corresponding band structure is shown in Fig.~\ref{fig1}(a). 
The Fermi surface is composed of spherical electron and hole pockets, respectively, while the degree of compensation can be controlled by changing the Fermi energy $E_{F}$ (Fig.~\ref{fig1}(b-d)). The system is perfectly compensated, which implies equal concentration $n_e=n_h$ of electrons and holes, when the Fermi energy $E_{F}=0$~eV (Fig.~\ref{fig1}(b)). This results in exactly $B^2$ dependence of MR (black solid line in Fig.~\ref{fig1}(e)) and zero Hall resistivity $\rho_H=0$ (black solid line in Fig.~\ref{fig1}(f)). This is fully consistent with the two-band model often used when 
interpreting experimental data if one further assumes equal mobilities $\mu_e=\mu_h$ of electrons and holes~\cite{pippardbook}: 
\begin{equation}
\rho_{xx}=\frac{1}{e}\frac{(n_e \mu_e +n_h \mu_h)+(n_e \mu_e \mu_h^2 +n_h \mu_h \mu_e^2)B^2}{(n_e \mu_e +n_h \mu_h)^2+(n_e-n_h)^2 \mu_e^2 \mu_h^2 B^2},
\label{eqn-2band-rhoxx}
\end{equation}
\begin{equation}
\rho_{xy}=\frac{1}{e}\frac{(n_h \mu_h^2 -n_e \mu_e^2)B+ \mu_e^2 \mu_h^2(n_h -n_e )B^3}{(n_e \mu_e +n_h \mu_h)^2+(n_e-n_h)^2 \mu_e^2 \mu_h^2 B^2}.
\label{eqn-2band-rhoxy}
\end{equation}
When the Fermi level is shifted up or down in energy, e.g. $E_\text{F}=0.2$~eV  and $E_\text{F}=-0.2$~eV in Figs.~\ref{fig1}(c) and \ref{fig1}(d), respectively, the perfect compensation is lifted in favor of a larger concentration of correspondingly electrons and holes. In this case, the field dependence of MR deviates from the quadratic behavior, as a power law of $B^{1.9}$ (dashed curves in Fig.~\ref{fig1}(e)),  and could even saturate at large field magnitudes depending on the degree of compensation. Meanwhile, the Hall resistivity is positive (negative) with nonlinear field dependence instead of zero as in the case of perfectly compensated system. This reflects the dominance of holes (electrons) in the conduction process. Given our Hamiltonian has particle-hole symmetry, the MR values in the two cases  are equal, while the Hall resistivity differs only in sign but not in magnitude. We note that due to the spherical shape of both Fermi surface pockets the resistivities do not depend on the magnetic field orientation.

\begin{figure}[t]
\begin{center}
\includegraphics[width=6cm]{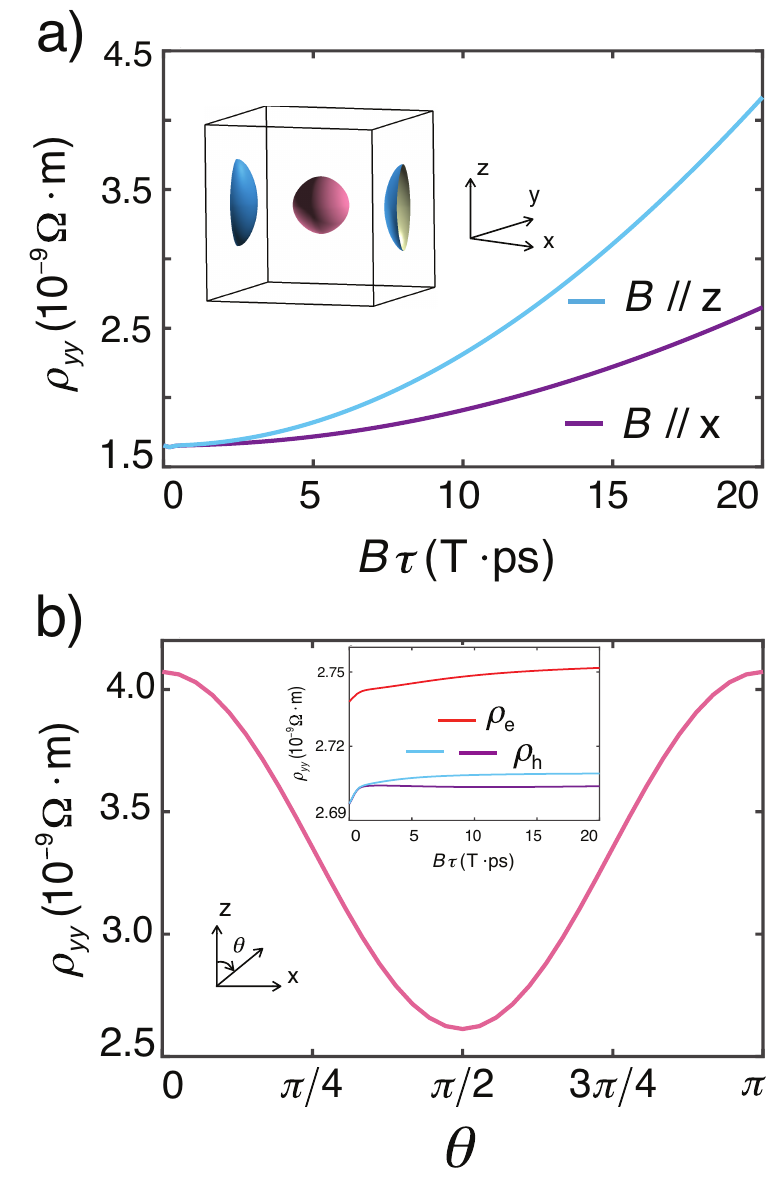}
\caption{(a) Field dependence of resistivity $\rho_{yy}$ in the anisotropic two-band model described by Hamiltonian (\ref{2band_aniso_H}). The inset shows the Fermi surface. (b) Resistivity $\rho_{yy}$ as a function of magnetic field orientation. The inset shows individual resistivities $\rho_{yy}$ of electrons (red) and holes (blue for $B$$\parallel$$z$ and purple for $B$$\parallel$$x$). The resistivity of electrons is scaled as ${\rho_{yy}}/{1.62}$ in order to make this curve visible.}
\label{fig2}
\end{center}
\end{figure}

\textit{Anisotropic two-band model.} As a next step towards more general model, we introduce a Fermi surface anisotropy considering the following Hamiltonian:
\begin{equation}
H(\bold{k})=d_0(k_x,k_y, k_z) \sigma_0 +d_z(k_x,k_y, k_z) \sigma_z ,
\label{2band_aniso_H}
\end{equation}
with $d_0 = [-3\cos(k_x)+\cos(k_y)-0.5\cos(k_z)-1.5]$ and $d_z = [\cos(k_x)-3\cos(k_y)- 1.5\cos(k_z) +6.5]$.
For $E_\text{F}=0$ the Fermi surface consists of a spherical electron pocket and an elliptical hole pocket, which has a longer axis along the $z$ direction, and two identical shorter axes in the $x$$-$$y$ plane (Fig.~\ref{fig2}(a), inset). The anisotropic shape of the hole Fermi surface results in different velocities and cyclotron masses under different magnetic field orientations, and hence affects the degree of compensation between electron and hole charge carriers. In this model, the contribution to the conductivity due to electrons does not change because of the isotropic Fermi surface, but that due to holes is sensitive to the direction of field $B$ as shown in Fig.~\ref{fig2}. The $\rho_{yy}$ resistivity plotted in Fig.~\ref{fig2}(a) is much larger for $B$$\parallel$$z$ compared to $B$$\parallel$$x$. This implies that magnetic field parallel to the $z$ axis favors more efficient charge-carrier compensation in this system. Fig.~\ref{fig2}(b) showing $\rho_{yy}$ as a function of magnetic field orientation, often referred to as the angular MR, reaffirms that the system attains its maximum resistivity for magnetic field along the $z$ direction. The degree of compensation is further assessed by plotting separately the resistivities of electron and hole charge carriers in Fig.~\ref{fig2}(b), inset.  
The fact that the resistivity of holes is larger for $B$$\parallel$$z$ is expected from the Drude model, where the conductivity is written as $\sigma=\frac{n e \mu}{1+ \mu^2 B^2}$. Considering the ellipsoid shape of the hole pocket (Fig.~\ref{fig2}(a), inset), the holes have a smaller cyclotron mass for $B$$\parallel$$z$ compared to $B$$\parallel$$x$, and hence a larger mobility since $\mu=\frac{e \tau}{m^*}$~\cite{supplementary}. The conductivity is smaller for $B$$\parallel$$z$ in contrast to $B$$\parallel$$x$, complying with $\sigma=\frac{n e \mu}{1+ \mu^2 B^2}$, while the opposite is true for the resistivity. Equivalently, one can also argue that the trajectories of charge carriers with larger mobility are more easily altered by the external magnetic field, which gives rise to a larger MR.

\begin{figure}[b]
\begin{center}
\includegraphics[width=5.7cm]{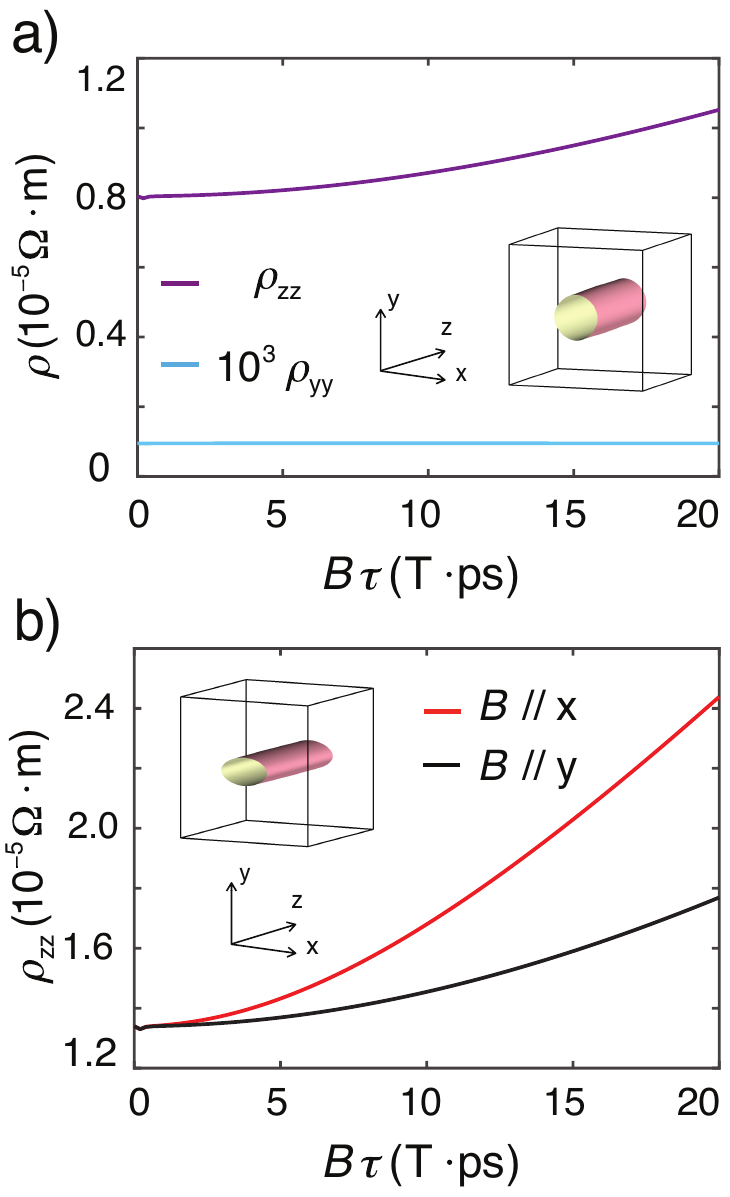}
\caption{(a) Field dependence of resistivities $\rho_{yy}$ and $\rho_{zz}$ for the isotropic open Fermi surface model. The magnetic field is applied along the $x$ axis. The $\rho_{yy}$ resistivity is multiplied by $10^3$ to make it visible.
(b) Field dependence of resistivity $\rho_{zz}$ for the anisotropic open Fermi surface model with magnetic field oriented along the $x$ and $y$ axes. The insets show the corresponding Fermi surfaces.}
\label{fig3}
\end{center}
\end{figure}

\textit{Open Fermi surface models.} The presence of open orbits is another physical mechanism responsible for non-saturating MR. The Fermi surface shown in Fig.~\ref{fig3}(a) inset represents an open cylinder along the $k_z$ direction described by Hamiltonian $H(\bold{k})=-2\cos k_x -2\cos k_y-\epsilon\cos k_z +3$. A small $k_z$ dispersion given by $\epsilon = 0.02$ is introduced in order to avoid numerical instabilities in our computations. Assuming $B$ is oriented along the $x$ axis, one can write a general form of the conductivity tensor~\cite{pippardbook}
\begin{align}
\begin{adjustbox}{max width=0.9\textwidth}
$\sigma_{ij} \approx c_0
\left(
\begin{array}{cccc}
c_1  &  c_2 &  -\frac{c_3}{\mu B} \\
c_2  &  c_4+\frac{c_5}{\mu^2 B^2} &  -\frac{c_6}{\mu B} \\
\frac{c_3}{\mu B}  & \frac{c_6}{\mu B} & \frac{c_7}{\mu^2 B^2}
\end{array}
\right)$ ,
\end{adjustbox}
\end{align}
from which the two elements of the resistivity tensor are $\rho_{yy}  \approx \frac{c_1c_7+c_3^2}{\mu^2 B^2 \vert \sigma \vert }$ and $\rho_{zz}  \approx \frac{c_1c_4-c_2^2}{\vert \sigma \vert}$, with constants $c_0$--$c_7$, mobility $\mu$, and determinant of the conductivity tensor $\vert \sigma \vert$. Since $\vert \sigma \vert$ has a leading term proportional to $\frac{1}{\mu^2 B^2 }$, $\rho_{zz}$ increases nearly quadratically with magnetic field $B$, while $\rho_{yy}$ saturates, as shown in Fig.~\ref{fig3}(a). Since open orbit extend along the $z$ axis in presence of field $B$ parallel to the $x$ axis, only the average velocity $v_y$ remains finite when $\omega \tau \gg 1$. Therefore, $\sigma_{yy}$ is a constant in contrast to the very small velocity $v_z$ resulting in significant resistivity $\rho_{zz}$. Non-saturating MR resulting from open orbits is observed 
for current along the open orbits and field $B$ applied in an orthogonal direction. This property can be used in order to distinguish the open-orbit mechanism of non-saturating MR from the charge-carrier compensation scenario. As illustrated in Fig.~\ref{fig3}(a), inset for $B$ oriented along the $x$ axis, the non-saturating MR is observed only when current is applied in the $z$ direction (purple line), while the  $\rho_{yy}$ resistance saturates (blue line).

To complete the discussion, we introduce a deformed, anisotropic open Fermi surface shown in the inset of Fig.~\ref{fig3}(b) and described by Hamiltonian $H(\bold{k})=-2\cos k_x-6\cos k_y-\epsilon\cos k_z+7$. The Fermi surface has no rotation symmetry in the $x$$-$$y$ plane, and hence the resistivity is expected to exhibit anisotropy upon changing the orientation of magnetic field. For $B$$\parallel$$y$, the flattened cylindrical Fermi surface gives rise to $\bold{v} \propto \nabla_{\bold{k}}\varepsilon(\bold{k})$ parallel to the field $B$, hence the Lorentz force is small for these charge carriers, and their contribution to MR is small as well. In contrast, for $B$$\parallel$$x$ the carriers have velocity normal to the field direction resulting in large resistivity, as shown in Fig.~\ref{fig3}(b). This model shows that the effective velocity of carriers changes under different magnetic field orientations giving rise to anisotropic resistivity, as we will see below when discussing magnetotransport in realistic materials.

\section{Representative materials}
\label{Results}
\subsection{Copper}

\begin{figure}
\begin{center}
\includegraphics[width=8.5cm]{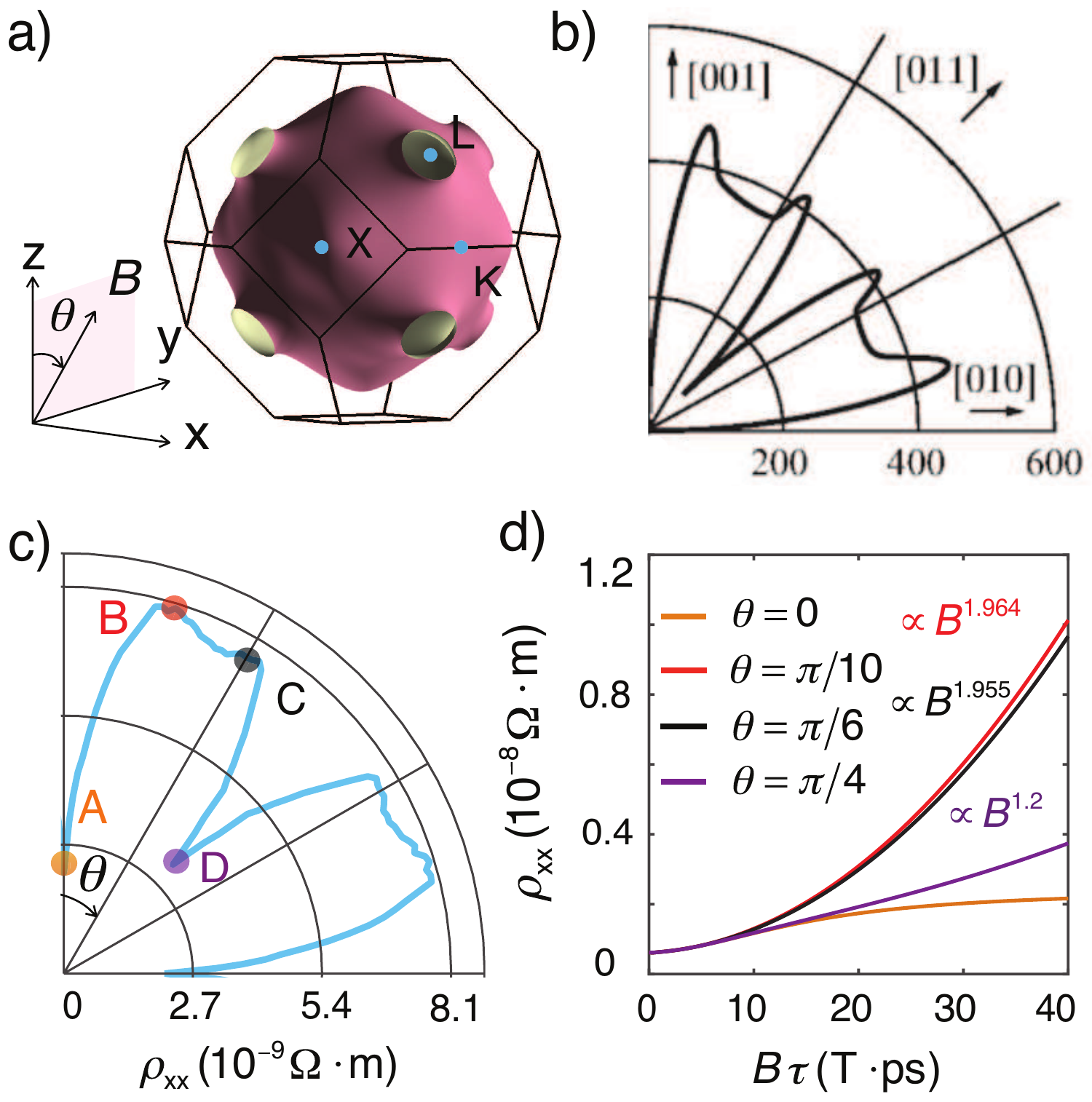}
\caption{(a) Fermi surface of copper.
(b) Polar diagram that shows the dependence of experimentally measured $\delta \rho_{xx}(B)/ \rho(0)$ on magnetic field direction (adopted from Ref.~\onlinecite{Cu-anisotropyMR-book}). The measurements were performed on a single crystal of copper at $T=4.2$~K and $B = 1.8$~T rotated in the plane normal to the current direction. 
(c) Calculated anisotropy of resistivity $\rho_{xx}$ for magnetic field rotated in the  $y$$-$$z$ plane agrees well with experimental results in panel (b) given $B\tau$ corresponds to $\omega \tau \gg 1$. 
(d). Resistivity $\rho_{xx}$ as a function of the magnitude of magnetic field $B$ for the four field directions indicated in panel (c).}
\label{fig4}
\end{center}
\end{figure}

Copper is a late $3d$ transition metal that crystallizes in the face-centered cubic lattice and has a relatively simple Fermi surface. The fully populated $d$ band of copper does not contribute to the Fermi surface, while the free-electron-like $s$ band is subject to a sufficiently strong periodic potential. With the minimum energy at the $\Gamma$ point, the free-electron-like band attains the Fermi energy  $E_\text{F}$  before reaching the Brillouin zone boundary along the $\Gamma$--X and $\Gamma$--K directions, but not along the $\Gamma$--L direction. This results in characteristic ``neck'' features in the Fermi surface, as shown in Fig.~\ref{fig4}(a), leading to rich magnetotransport properties. 
Fig.~\ref{fig4}(b) reproduces the experimental polar diagram of angular MR measured for a single crystal of copper at $T=4.2$~K in magnetic field $B=1.8$~T rotated in the $y$$-$$z$ plane~\cite{Cu-anisotropyMR-book}. It was found that MR increases quadratically with increasing magnetic field for most field orientations, while for others it saturates quickly~\cite{Cu-anisotropyMR-book}. The calculated  angular MR (Fig.~\ref{fig4}(c)) shows excellent agreement with the experimental results reproducing all qualitative features. 

We will now focus on discussing the resistivity anisotropy in connection with the Fermi surface topology. Firstly, resistivity anisotropy reflects the symmetry of the Fermi surface projected  onto the plane perpendicular to current. The Fermi surface of copper has cubic symmetry, i.e. $\rho(\theta)=\rho(\theta+{\pi}/{2})$, and thus only a quarter of the polar diagram is shown. Furthermore, considering the periodicity of reciprocal space a variety of orbits is expected for different  magnetic field orientations, thus leading to complex pattern of the anisotropic magnetoresistivity. As shown in Figs.~\ref{fig4}(b) and \ref{fig4}(c), the peaks in resistivity correspond to intermediate angles rather than high-symmetry orientations of magnetic field. The resistivity grows quickly from a minimum at $\theta=0$ to a maximum at approximately $\theta = {\pi}/{10}$, drops slightly to form a plateau, shows another peak close to $\theta = {\pi}/{6}$, and then decreases rapidly to another resistivity minimum at $\theta={\pi}/{4}$. The behavior of resistivity between $\theta={\pi}/{4}$ to  $\theta={\pi}/{2}$ shows the same features but in reverse order owing to the cubic symmetry of the Fermi surface. 

\begin{figure}
\begin{center}
\includegraphics[width=7.8cm]{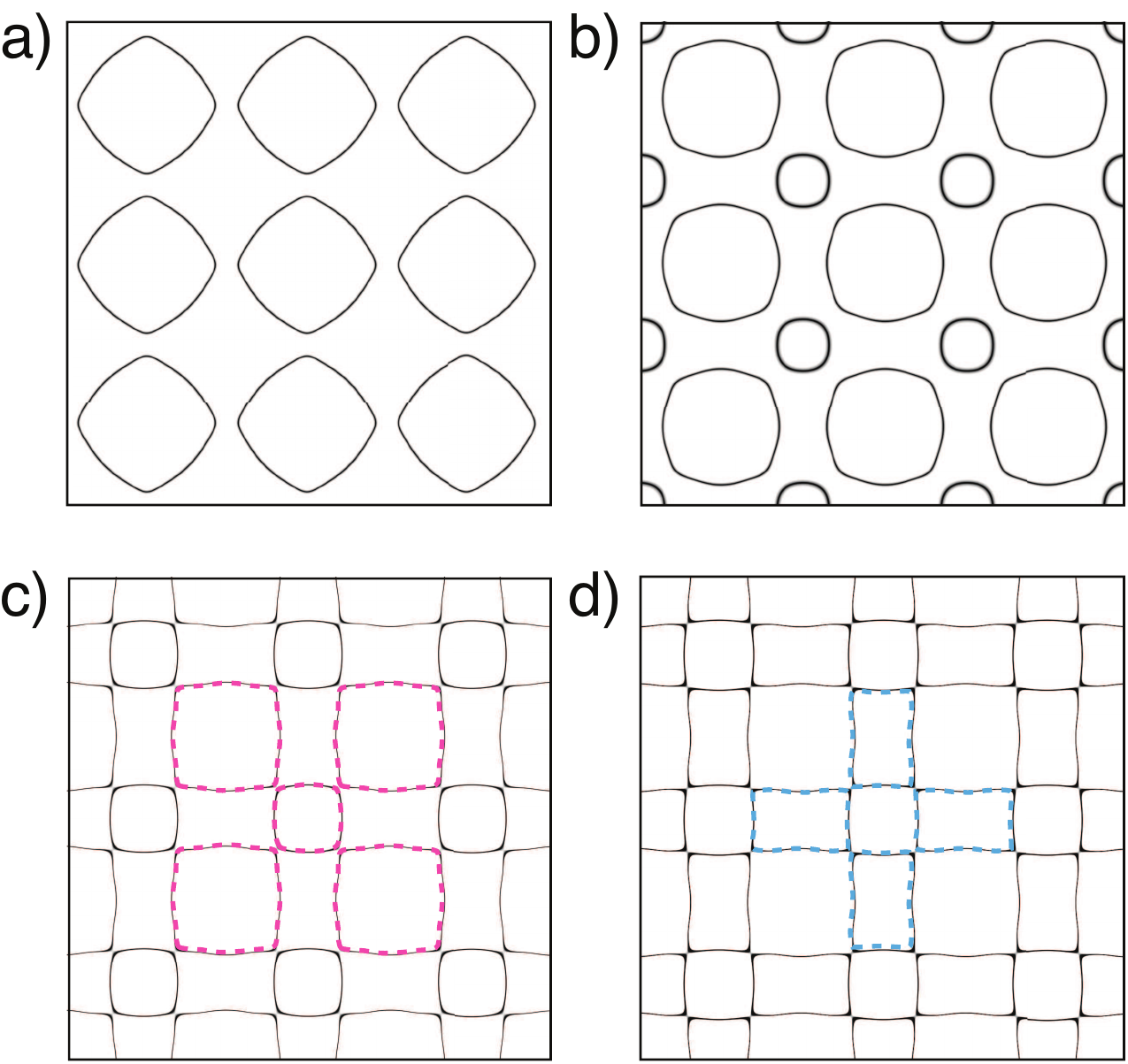}
\caption{Typical cross-sections of the Fermi surface of copper for magnetic field oriented along the $z$ axis (point A in Fig.~\ref{fig4}(c)). 
The horizontal axis is along the $k_x$ direction. Cross-sections in the $k_x-k_y$ plane correspond to (a) $k_z=0$, (b) $k_z={0.3\pi}/{a}$, (c) $k_z={0.62\pi}/{a}$, and (d) $k_z={0.63\pi}/{a}$.
Pink and blue dashed lines highlight the closed electron and hole orbits, respectively.}
\label{fig5}
\end{center}
\end{figure}

\begin{figure}
\begin{center}
\includegraphics[width=8.6cm]{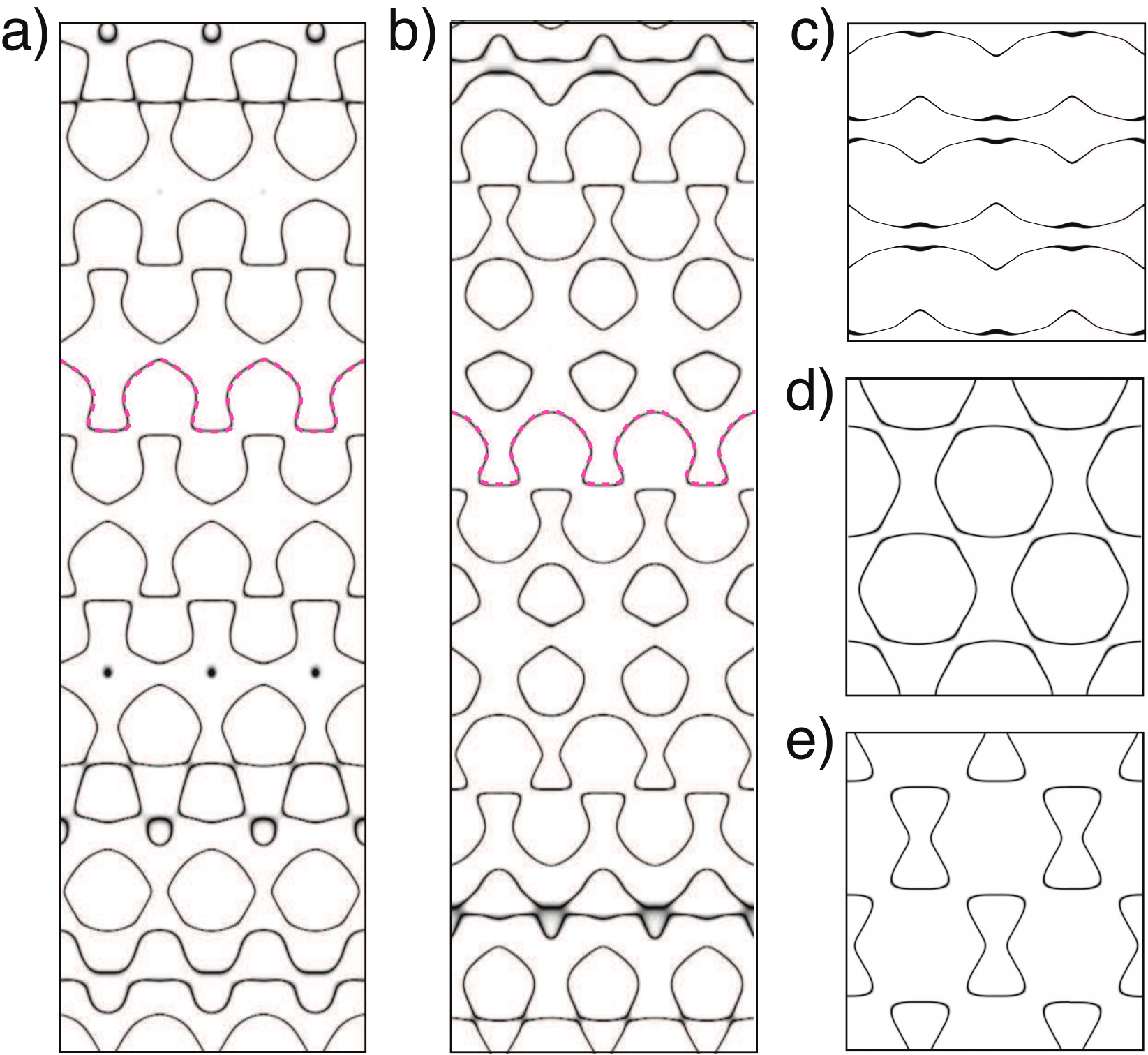}
\caption{Typical cross-sections of the Fermi surface of copper that correspond to magnetic field orientations marked by points (a) B, (b) C and (c)-(e) D in Fig.~\ref{fig4}(c).
The horizontal axis corresponds to the $k_x$ direction while the vertical axis is the direction parallel to $\hat{k}_x \times {\bf B}$.  In panels (a)-(c) the plane includes the $\Gamma$ point, while those in panels  (d) and (e) pass through points $(0,{0.15\pi}/{a},{0.15\pi}/{a})$  and $(0,{0.52\pi}/{a},{0.52\pi}/{a})$, respectively. 
The pink dashed lines highlights one of the open orbits in (a) and (b). Panel (c) shows open orbits extend along the $k_x$ direction. Panels (d) and (e) show closed electron and hole orbits, respectively.  }
\label{fig6}
\end{center}
\end{figure}

In order to understand the physics underlying the magnetotransport anisotropy, in Fig.~\ref{fig4}(d) we plot the calculated field dependence of resistivity for magnetic field orientations that correspond to extrema points marked by A, B, C and D in Fig.~\ref{fig4}(c). The corresponding representative orbits realized at point A are summarized in Fig.~\ref{fig5}, while typical scenarios at field orientations B, C and D are presented in Fig.~\ref{fig6}.

For magnetic field oriented along the $z$ axis (point A) there are two distinct ways a plane normal to it can cut the Fermi surface: either crossing or not the ``necks'' of the Fermi surface. It is obvious that when the plane does not cross the ``necks'' simple closed electron orbits are formed. Such orbits shown in Figs.~\ref{fig5}(a) and \ref{fig5}(b) for the planes defined by $k_z=0$ and $k_z={0.3 \pi}/{a}$, respectively, appear as circles deformed by the effect of the periodic potential. Figs.~\ref{fig5}(c) and \ref{fig5}(d) shows the cross-sections produced by planes defined by $k_z={0.62 \pi}/{a}$ and $k_z={0.63 \pi}/{a}$, respectively, and the resulting orbits are highlighted in pink and blue. The $k_z={0.62 \pi}/{a}$ plane almost crosses the ``necks'' representing the extreme case of orbits shown in  Fig.~\ref{fig5}(b).
In contrast, the $k_z={0.63 \pi}/{a}$ plane crosses the ``necks'' and the fragments of orbits on the Fermi surface in the adjacent periodic replicas of the Brillouin zone join to form closed hole orbits as these orbits enclose empty states (Fig.~\ref{fig5}(d)). Assuming that charge carriers in orbits shown in Figs.~\ref{fig5}(a)--\ref{fig5}(c) move clockwise, those in Fig.~\ref{fig5}(d) move in the opposite direction. Consequently, magnetoresistance saturates at high fields due to incomplete compensation of these two kinds of charge carriers at this particular direction of magnetic field. 

Upon tilting magnetic field away from the $z$ axis in the $y$$-$$z$ plane open orbits extending along the $k_x$ direction emerge. Therefore, the resistivity $\rho_{xx}$ increases and tends to show non-saturating behavior (field orientations between points A and B in Fig.~\ref{fig4}(c)). At point B ($\theta={\pi}/{10}$), the resistance achieves its maximum, and shows a nearly quadratic $B^{1.964}$ magnetic field scaling with no saturation (Fig.~\ref{fig4}(d)). In order to confirm that this behavior originates from open orbits we plot a typical Fermi surface cross-section (Fig.~\ref{fig6}(a)). One can observe a series of open orbits extending along the $k_x$ direction with one of them highlighted in pink. As explained above, open orbits along $k_x$ would result in few charge carriers with velocity in this direction, and hence in parabolic dependence of resistance upon increasing magnetic field. However, the resistivity tends to depart from ideal parabolic scaling ($\rho_{xx} \propto  B^{1.964}$) due to the contribution of closed orbits seen in Fig.~\ref{fig6}(a).
At point C ($\theta={\pi}/{6}$), the resistivity $\rho_{xx}$ shows a similar $B^{1.955}$ field dependence (Fig.~\ref{fig4}(d)). A typical typical Fermi surface cross-section shown in Fig.~\ref{fig6}(b) reveal the dominance of open orbits extending along the $k_x$ direction, while the presence of closed orbits can also be noted which justifies the observed sub-quadratic dependence of resistivity on magnetic field.

As the field orientation changes past point C, the resistivity declines quickly reaching its minimum at point D ($\theta={\pi}/{4}$), and this minimum is distinct from the one at point A ($\theta=0$ and $\theta={\pi}/{2}$ by symmetry). The resistivity does not show quadratic scaling but rather grows as $B^{1.2}$ without any sign of saturation in contrast to the field dependence at point A. Three typical Fermi surface cross-sections shown in Figs.~\ref{fig6}(c)--\ref{fig6}(e) demonstrate that this field orientation gives rise to a more complicated situation in which open orbits extending along the $k_x$ direction (Fig.~\ref{fig5}(c)) as well as compensation of electrons (Fig.~\ref{fig6}(d)) and holes (Fig.~\ref{fig6}(e)) are both present. 

This example shows how the Boltzmann approach calculations help understanding the physical mechanism underlying transverse MR in copper, a prototypical nearly free-electron metal. Not only these calculations reproduce the experimentally observed delicate features in the dependence of magnetotransport on the orientation of magnetic field, but also allow interpreting these features in terms of the interplay between the open-orbit and compensation mechanisms, and hence the Fermi surface topology.


\subsection{Bismuth}

\begin{figure*}[!t]
\begin{center}
\includegraphics[width=14cm]{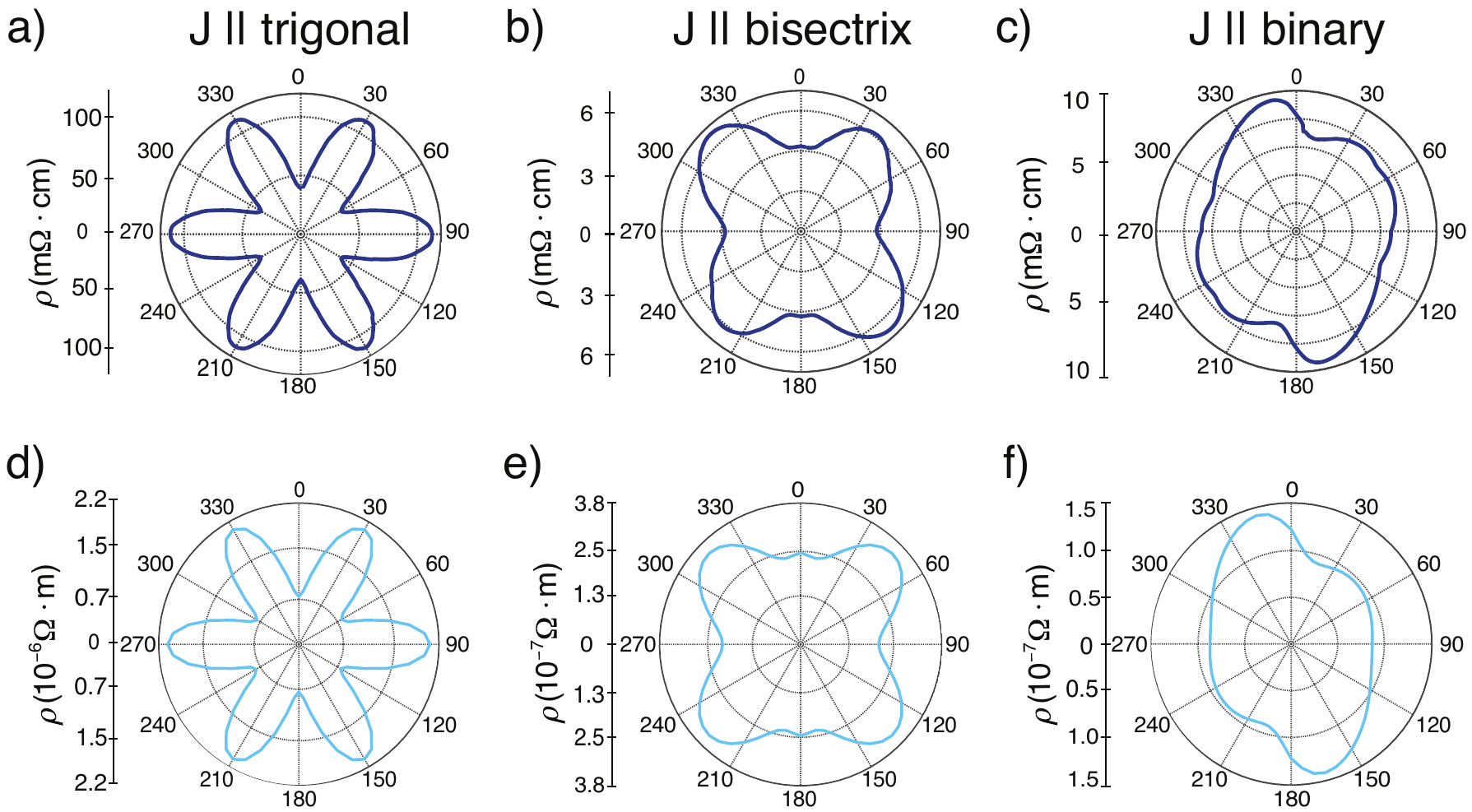}
\caption{Comparison of the experimentally measured and calculated resistivity anisotropy of bismuth. (a)-(c) Resistivity anisotropy of bismuth measured at magnetic field $B = 0.5$~T and temperatures $T = 10$~K, $T = 20$~K and $T = 15$~K, respectively, for the three indicated current directions (reproduces Fig.~5 in Ref.~\onlinecite{bismuth-PRX.5.021022}). (d)-(f) Calculated resistivity anisotropy of bismuth for the same current directions. We assumed $B\tau = 2.2$~T$\cdot$ps in our calculations, and similar results can be obtained at weaker magnetic fields.}
\label{fig7}
\end{center}
\end{figure*}

Bismuth is perhaps the most extensively studied material that shows extremely large non-saturating MR reaching $1.6 \times 10^7\%$ at $T=4.2$~K and $B=5$~T~\cite{bismuth-PhysRev.91.1060}. Furthermore, its MR exhibits a very strong dependence on the magnetic field orientation that can be observed even at room temperature and fields as low as $B=0.7$~T~\cite{bismuth-Zhu2011}, which is in sharp contrast to its almost isotropic electrical conductivity in zero applied magnetic field. The resistivity anisotropy in presence of magnetic field is believed to stem from its peculiar semimetallic Fermi surface, which consists of one small ellipsoid hole pocket located at the $T$ point and three small ellipsoid electron pockets located at the $L$ point.

Previously, Aubrey described the magnetoconductivity tensor of bismuth as $\hat{\sigma}(B)=n e (\hat{\mu}^{-1}+\hat{B})^{-1}$, where $\hat{\mu}$ and $\hat{B}$ are the effective mobility and magnetic field tensors, respectively~\cite{Aubrey-0305-4608-1-4-321}. Within this formalism, angular MR showed reasonable agreement with experimental results assuming appropriate values for the components of the mobility tensor for the electron and hole charge carriers. Here, we employ the tight-binding Hamiltonian obtained from DFT calculations without assuming any parameters to obtain field-dependent resistivities for different current directions shown in Fig.~\ref{fig7}. The upper panels of Fig.~\ref{fig7} reproduce experimental results from Ref.~\onlinecite{bismuth-PRX.5.021022}, while the lower panels are the results of  our calculations of angular MR. Further details can be found in the Supplemental Material~\cite{supplementary}.

\begin{figure}
\begin{center}
\includegraphics[width=8.6cm]{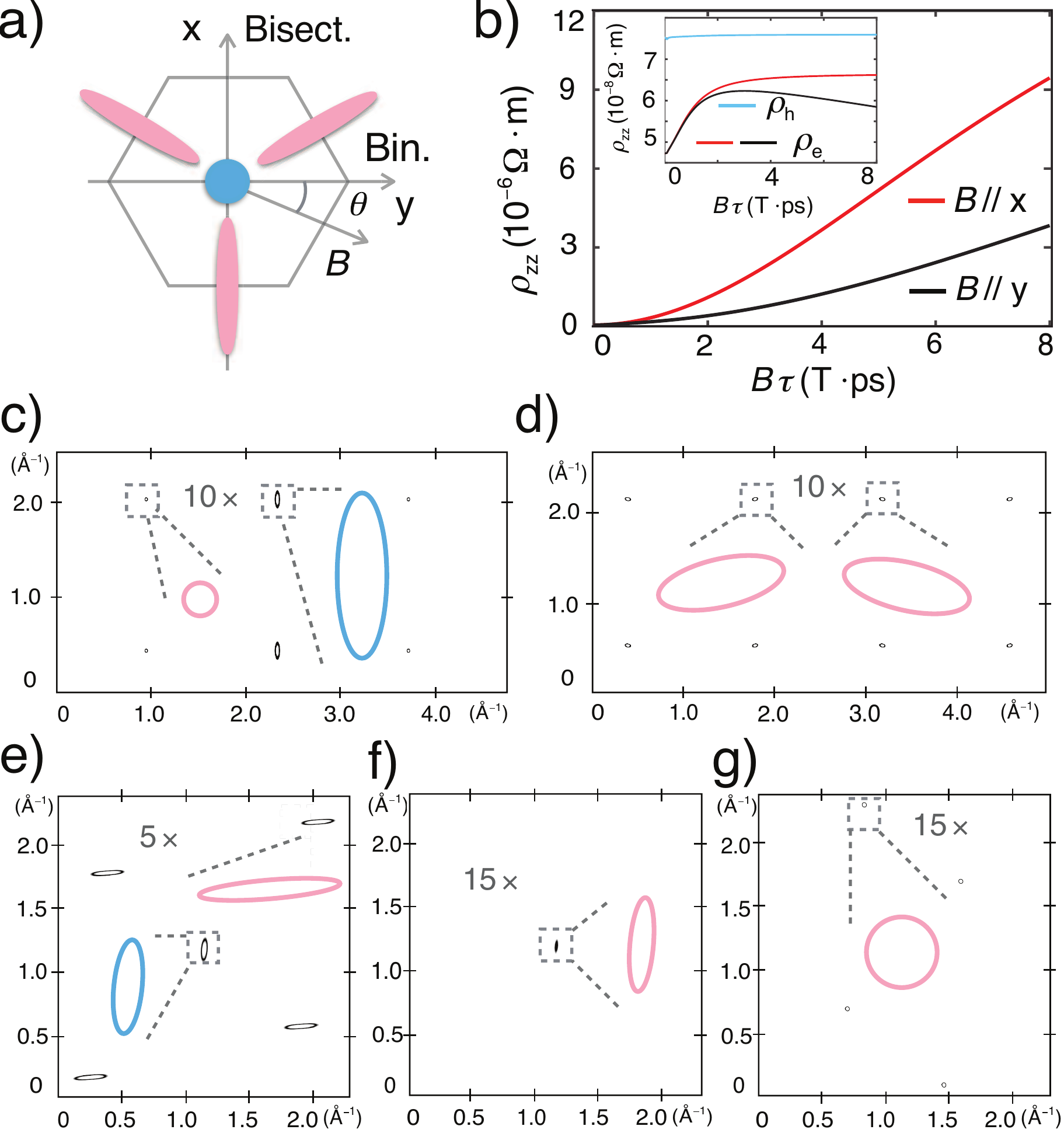}
\caption{(a) Schematic drawing of the Fermi surface projection of bismuth onto the $x$$-$$y$ plane normal to the trigonal axis. Hole and electron pockets are shown in blue and pink, respectively. (b) Magnetic field dependence of $\rho_{zz}$ for magnetic field $B$ oriented along the bisectrix and binary axes. The inset shows $\rho_{zz}$ for electron and hole contributions separately, with the resistivity of hole charge carriers scaled to $\rho_{zz}/{22}$ in order to fit the plot. (c),(d) Cross-sections of the Fermi surface for the case of magnetic field $B$ oriented along the $x$ axis. Horizontal and vertical axes are along the $k_y$ and $k_z$ directions, respectively. (e)-(g) Cross-sections of the Fermi surface for magnetic field $B$ along the $y$ axis.  Horizontal and vertical axes are along the $k_x$ and $k_z$ directions, respectively. Enlarged cross-sections are shown due to the very small size of the Fermi surface pockets in bismuth.}
\label{fig8}
\end{center}
\end{figure}

Fig.~\ref{fig7} presents angular MR for current applied along the three high-symmetry directions while magnetic field is rotated in the plane normal to the current. We will first analyze the symmetry of the MR anisotropy curves. For current oriented along the trigonal axis (below referred to as the $z$ axis), the resistance shows a six-fold symmetry (Figs.~\ref{fig7}(a) and \ref{fig7}(d)). This is a consequence of the symmetry of the Fermi surface projected onto the plane perpendicular to current (see Fig.~\ref{fig8}(a) for schematic illustration of the Fermi surface pockets). In this case, the Fermi surface projection has $C_3$ symmetry, therefore resistivity is invariant under rotation of the field by ${2\pi}/{3}$. Furthermore, inversion symmetry $\rho_{zz}(\theta)=\rho_{zz}(-\theta)$ results in six-fold rotational invariance of the resistivity. When current is applied along the bisectrix (the $x$ axis), the configuration of the projected Fermi surface has only mirror and inversion symmetries, therefore  resistance complies with $\rho_{xx}(\theta)=\rho_{xx}(-\theta)$ and $\rho_{xx}(\theta)=\rho_{xx}(\pi + \theta)$ (Figs.~\ref{fig7}(b) and \ref{fig7}(e)). There remains only inversion symmetry in the case of configuration of the projected Fermi surface for the current applied along the binary axis (the $y$ axis), resulting in $\rho_{yy}(\theta)=\rho_{yy}(\pi + \theta)$ (Figs.~\ref{fig7}(c) and \ref{fig7}(f)).

We will discuss in detail only the first configuration as an example in order to demonstrate the origin of angular MR. Since there are no open orbits in the Fermi surface of bismuth, it is the variation of the degree of compensation of the two types of charge carriers that make the resistivity change upon rotation of magnetic field. From the sketch of configuration in Fig.~\ref{fig8}(a), changing the orientation of magnetic field does not affect the hole pocket, and thus the concentration and mobility of the hole carriers are constant. Therefore, the resistivity of holes would saturate upon increasing the strength of magnetic field, as shown by the blue curve in Fig.~\ref{fig8}(b), inset. In contrast, the three electron pockets are highly anisotropic, hence rotating the direction of magnetic field would alter both the concentration and mobility of the electron carriers. Following the analysis described above, electrons would obtain a larger effective resistivity when magnetic field is oriented along the long axis of the corresponding Fermi surface pockets. Consequently, the compensation should be more efficient for $B$$\parallel$$x$ than $B$$\parallel$$y$. This is confirmed in Fig.~\ref{fig8}(b), which shows that the resistivity is much larger for  $B$$\parallel$$x$ than $B$$\parallel$$y$. In addition, in Fig.~\ref{fig8}(b), inset, we plot separately the individual resistivities of holes and electrons. Combined with the  symmetry analysis given above, we can understand all features of the angular MR polar plot in Figs.~\ref{fig7}(a) and \ref{fig7}(d).

In order to complete the discussion, typical cross-sections of the Fermi surface for $B$$\parallel$$x$ and $B$$\parallel$$y$ are drawn in Figs.~\ref{fig8}(c)-(d) and Figs.~\ref{fig8}(e)-(f), respectively. These plots show that the orbits for the electron and hole charge carriers are quite different for these two field orientations in both shape and size. Moreover, it is the multi-pocket (valley) anisotropy that makes the saturation of resistivity for electrons require a larger magnetic field strength compared to that of the holes, as show in the inset of Fig.~\ref{fig8}(b). Following the steps outlined in the case of current parallel to the trigonal axis, one can arrive to the conclusion that resistivity increases when the carriers are more effectively compensated.

\subsection{Type-II Weyl semimetal WP$_2$}

The $\beta$-phase of WP$_2$ is a recently predicted type-II Weyl semimetal in which the neighboring Weyl points have the same chirality, thus making the topological phase stable against small lattice perturbations~\cite{WP2-PRL.117.066402}. Soon after the prediction, experiments have shown that single crystals of WP$_2$ exhibit extremely large MR of $4.2\times 10^{6}\%$ at $T=2$~K and $B=9$~T, and reaching over $2 \times 10^8\%$ at $T=2.5$~K and $B=63$~T~\cite{WP2-Kumar2017}. Moreover, the observed transverse MR showed a high degree of anisotropy, i.e. MR for the magnetic field along the $b$ and $c$ axes differ by more than two orders of magnitude, much higher than in another candidate type-II Weyl semimetal WTe$_2$~\cite{MRWTe2-Ali2014}. The observed MR anisotropy was attributed to open orbits under the field oriented along the $c$ axis and the anisotropic shape of the hole pocket~\cite{WP2-Kumar2017}.

We have chosen WP$_2$ as another representative example of semimetal showing XMR that is nevertheless much less understood. In contrast to bismuth discussed above, the Fermi surface of WP$_2$ is much more extended and has a complex shape. Even being located relatively close to the Fermi level, the Weyl points of opposite chirality are enclosed within the same electron pocket~\cite{WP2-PRL.117.066402}. Therefore, in undoped WP$_2$ all Fermi surface sheets are topologically trivial, that is have zero Chern number. The transport properties specific to the Weyl fermion quasiparticles thus do not manifest in undoped WP$_2$, and the employed semiclassical treatment can be used without explicitly introducing the Berry curvature.

\begin{figure*}
\begin{center}
\includegraphics[width=15cm]{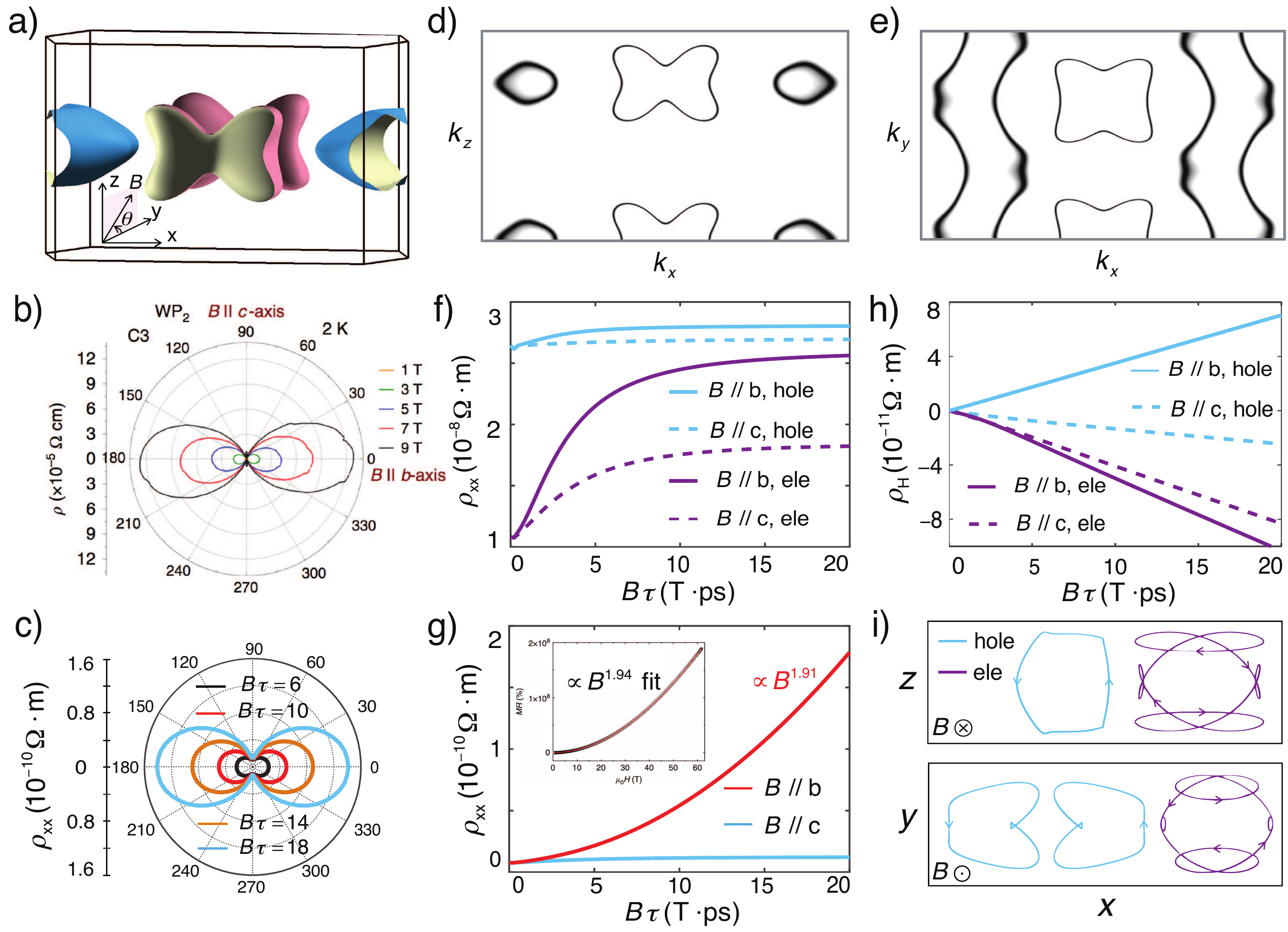}
\caption{(a) Fermi surface of WP$_2$ composed of a bowtie electron pocket (pink) and an open tube hole pocket (blue) extending along the $y$ direction. (b) Measured (adopted from Ref.~\onlinecite{WP2-Kumar2017}) and (c) calculated resistivity $\rho_{xx}$ as a function of magnetic field orientation for different magnetic field strengths, with $B$$\parallel$$b$ being the reference field orientation. (d),(e) Fermi surface cross-sections for magnetic field along the $b$ and $c$ axes, respectively. (f) Individual resistivities $\rho_{xx}$ of electron and hole charge carriers for $B$$\parallel$$b$ and $B$$\parallel$$c$. (g) Field dependence of resistivity $\rho_{xx}$ for the cases of magnetic field oriented along the $b$ and $c$ directions. The inset shows experimentally measured $\rho_{xx} \propto B^{1.94}$ for $B$$\parallel$$b$ that can be compared with the calculated $\rho_{xx} \propto B^{1.91}$ scaling (red line in the main figure (g)). (h) Individual Hall resistivities $\rho_{H}$ for electron and hole charge carriers in magnetic field oriented along the $b$ and $c$ directions. (i) Electron and hole orbits in real space for the Fermi surface cross-sections shown in panels (d) and (e). The orbits self-intersect due to the presence of concave segments of the Fermi surface~\cite{Ong-PRB.43.193}. }
\label{fig9}
\end{center}
\end{figure*}

The Fermi surface of WP$_2$ is spin-split due to spin-orbit coupling (SOC) in absence of inversion symmetry, but the effect of SOC on the Fermi surface topology is very weak. Below, we present the results of calculations carried out without SOC taken into account, while the results with SOC are discussed in the Supplemental Material~\cite{supplementary}. The calculated Fermi surface of WP$_2$ is composed of a bowtie-like closed electron pocket and a tube-shaped open hole pocket extending along the $a$ axis (Fig.~\ref{fig9}(a)), in agreement with previous calculations \cite{WP2-PRL.117.066402} and experiments~\cite{WP2-Kumar2017}. From the previous analysis, non-saturating MR due to open orbits can be observed for current applied along the $b$ axis, i.e. the direction in which the open orbits extend. However, as-grown crystals of WP$_2$ are needle-shaped with longer dimension aligned along the $a$ axis, which makes it difficult to apply current along shorter axes $b$ and $c$. Consequently, in experiments reported to date the resistivity is measured for current applied along the $a$ axis while magnetic field is in the $b$$-$$c$ plane.

Comparison of the resistivity anisotropy measured by Kumar~{\it et al.}~\cite{WP2-Kumar2017}~ (Fig.~\ref{fig9}(b)) and calculated by us (Fig.~\ref{fig9}(c)) shows excellent agreement. The MR anisotropy exhibits two-fold symmetry due to the $m_{xz}$ mirror plane symmetry. The MR achieves its maximum value for magnetic field oriented along the $b$ axis, then decrease very quickly as magnetic field is rotated away from $b$, and assumes its minimum value when magnetic field is along the $c$ axis. As illustrated in Section~\ref{model}, non-saturating MR resulting from open orbits can be ruled out unless current is applied along the $b$ axis. How to understand this highly anisotropic MR when current is applied along the $a$ axis within the charge-carrier compensation picture?

As we have seen above, the compensation of charge-carriers depends sensitively on the orientation of magnetic field when the Fermi surface departs from the free-electron spherical shape~\cite{Ong-PRB.43.193}. In order to investigate these effects, we draw the cross-sections of the Fermi surface and calculate the resistivity for magnetic field oriented along the $b$ and $c$ axes (Figs.~\ref{fig9}(d) and \ref{fig9}(e)).
At first glance, it seems that the orbits of electrons are similar for the two field orientations that correspond to closed bowtie-shaped Fermi surface cross-sections. This is in contrast to the hole Fermi surface cross-sections that are closed for $B$$\parallel$$b$ and open for $B$$\parallel$$c$. An immediate conclusion would be that the hole charge carriers are responsible for the observed angular MR of WP$_2$. However, a carefully analysis of the compensation effects reveals the opposite -- the resistivity anisotropy for electrons is much more pronounced as can be clearly seen in Fig.~\ref{fig9}(f). Furthermore, the charge-carrier compensation for $B$$\parallel$$b$ is indeed much more efficient as compared to the case of $B$$\parallel$$c$ (blue solid line approaches purple solid line in Fig.~\ref{fig9}(f)). 

As far as the field dependence of resistivity is concerned, our calculations also show good agreement with the experimental results (Fig.~\ref{fig9}(g)). The measured MR exhibits a nearly quadratic MR~$\propto B^{1.94}$ field dependence for $B$$\parallel$$b$, while the calculated resistivity scales as $\rho_{xx} \propto B^{1.91}$. This demonstrates a nearly perfect charge-carrier compensation for $B$$\parallel$$b$ magnetic field orientation. 
In contrast, the $B$$\parallel$$c$ resistivity saturates quickly with the increase of magnetic field strength, thus also confirming the observed MR anisotropy. Kumar~{\it et al.}~\cite{WP2-Kumar2017} reported that MR for magnetic fields oriented along the $b$ and $c$ axes differ by 2.5 order of magnitude. In our calculations, assuming the relaxation time of $\tau = 10^{-9}$~s~\cite{WP2-Kumar2017}, the resistivity for $B$$\parallel$$b$ is about 3 order of magnitude larger than that for $B$$\parallel$$c$ at field strength of several Tesla, which agrees well with the experiments. It is nevertheless difficult to explain such a large anisotropy only by comparing the individual resistivities of charge carriers, which are of the same order of magnitude. 

In order to gain additional insight, we investigated the calculated Hall resistivity $\rho_H$ shown as a function of magnetic field in Fig.~\ref{fig9}(h). For magnetic field oriented along the $b$ axis, the Hall resistivity is positive for holes and negative for electrons, as one would expect. Surprisingly, for $B$$\parallel$$c$ both electrons and holes show negative $\rho_H$, which suggests that charge carriers originating from the hole pocket effectively behave like electrons. Considering the Fermi surface of WP$_2$ has a complex shape that includes concave segments, the Hall resistivity can show non-trivial behavior as illustrated in Ref.~\onlinecite{Ong-PRB.43.193}. 
To confirm this, in Fig.~\ref{fig9}(i) we plot the orbits in real space that correspond to the Fermi surface cross-sections shown in Figs.~\ref{fig9}(d) and \ref{fig9}(e). Fig.~\ref{fig9}(i), upper panel, shows that the motion of electron and hole carriers takes place in clockwise and anti-clockwise senses, respectively. For magnetic field oriented along the $b$ axis charge carriers show the expected behavior, as they originate from the electron and hole pockets, thus resulting in their effective compensation and nearly quadratic non-saturating MR. In contrast, for magnetic field applied along the $c$ axis, both types of charge carriers show electron-like orbits in real space (Fig.~\ref{fig9}(i), lower panel), thus precluding efficient charge-carrier compensation and resulting in rapid saturation of MR.

We conclude that the origin of high MR anisotropy in WP$_2$ is the distinct type of charge-carrier compensation rather than open orbits of hole carriers. The degree of compensation changes drastically under different magnetic field orientations due to the peculiar geometry of the Fermi surface, especially that of the hole pocket. Overall, the topology of the Fermi surface plays a crucial role in explaining the large MR anisotropy in WP$_2$.


\section{Concluding discussion}
\label{Summary}
We have demonstrated how our Boltzmann transport theory calculations reproduce the experimentally observed MR and allow interpreting the finest features of its dependence on the magnetic field orientation. Even though we assumed a constant relaxation time $\tau$ and plotted all results as a function of $B\tau$, the relaxation time itself is an important parameter that deserves a dedicated discussion. At very low temperatures the mean free path of charge carriers tends to become isotropic since scattering is dominated by the effect of impurities, while the average distance between impurities is independent of direction~\cite{anisotropy-tau-PR.121.1320, anisotropy-tau-book}. In single crystal of cooper at $T=4.2$~K the relaxation time $\tau$ is in the range $0.4 \times 10^{-10} - 4 \times 10^{-10}$~s~\cite{Cu-tau-PR.129.1990}.
For bismuth, the experimentally measured $\tau$ are around $2.5 \times 10^{-10} - 6.0 \times 10^{-10}$~s~\cite{PhysRevLett.22.26}. With increasing temperature, the phonon scattering becomes highly anisotropic close to $0.2\Theta_{\text{D}}$, as illustrated in Ref.~\cite{Ong-PRB.43.193}, while the Debye temperature of bismuth $\Theta_{\text{D}}=100$~K~\cite{bismuth-tau-book}. In our work, we compare the results of calculations with experiment data at temperatures of at most $T=20$~K, while the discussion of the anisotropy of $\tau$ at higher temperatures is included in the Supplemental Material~\cite{supplementary}. In the case of WP$_2$, the estimated relaxation time in the experiments is $3.8 \times 10^{-9}$~s, while the quantum lifetime obtained from broadening of the Shubnikov–de Haas oscillations is $7.9 \times 10^{-13}$ s~\cite{WP2-Kumar2017}, indicating that it is safe to discuss magnetotransport in this material within the semiclassical approximation. In addition, by making correspondence between the  calculated and measured magnetoresistivities we obtain the magnitudes of relaxation time of $1.0\times10^{-10}$~s for copper, $9.5\times10^{-10}$~s for bismuth and $1.0\times10^{-9}$~s for WP$_2$, all comparable to the reported experimentally measured values.

To summarize, our detailed numerical investigation of transverse magnetoresistance and its anisotropy allows us to conclude that these properties can be well understood by considering the topology of the Fermi surface, provided the latter is correctly described. In copper, a simple nearly free-electron metal with open Fermi surface geometry, both compensation and open-orbit mechanisms contribute to magnetotransport resulting in an intricate angular MR diagram. In the case of bismuth, the complex compensation between multi-valley Fermi surface gives rise to a distinct MR anisotropy pattern that is also well reproduced by our calculations. Finally, for the recently discovered type-II Weyl semimetal WP$_2$ we find that a novel charge-carrier compensation mechnism rather than the presence of open-orbits is responsible for the observed strong and highly anisotropic MR. We believe our study provides guidelines to clarifying the physical mechanisms underlying the magnetoransport properties in a broad range of materials, and will allow addressing the role of the topological protection in the pronounced MR response observed in both topologically trivial and non-trivial materials.

\section{acknowledgments}
We would like to thank N. Kumar for granting us the persmission to reproduce their experimental figures and K. Behnia for sharing their experimental data. 
We acknowledge support by the NCCR Marvel. 
Y.L. was supported by the National Natural Science Foundation of China (Grant No. 61604013) and the Fundamental Research Funds for the Central Universities (Grant No. 2016NT10).
First-principles and transport calculations have been performed at the Swiss National Supercomputing Centre (CSCS) under Project No. s832 and the facilities of Scientific IT and Application Support Center of EPFL.








%

\end{document}